# Negative Magnetoresistance in Granular Bi - HTSC with Trapped Magnetic Fields

A. A. Sukhanov, V. I. Omelchenko

Institute of Radioengineering and Electronics of RAS, 141190, Fryazino, Vvedenskogo sq. 1, Russia

e-mail: sukh@ms.ire.rssi.ru

**Abstract**

Magnetoresistive properties of granular Bi-based HTSC with trapped magnetic fields are investigated in the temperature region near superconducting transition . The effect of trapped field and transport current values and orientations on the field dependence of magnetoresistance is studied. It is found that for the magnetic field parallel and the current perpendicular to trapping inducing field the field dependence of magnetoresistance is nonmonotonic and magnetoresistance turns out to be negative for small fields. The magnetoresistance sign inversion field increases roughly linear with the trapped magnetic field and slightly decrease with transport current. The results are explained in the framework of model of magnetic flux trapping in granules or superconducting loops embedded in weak links matrix.

## I. Introduction.

The magnetic flux trapping (MFT) in high temperature superconductors (HTSC) essentially changes their magnetic properties (magnetic moment, susceptibility, screening, magnetic field penetration and so on) [1, 2] as well as their transport properties, in particular resulting in the frozen magnetoresistance phenomenon near resistive transition region [3, 4].

Thus the investigation of the properties of HTSC with trapped magnetic fields (TMF) is interesting in itself and also is useful for understanding of MFT nature, particularly, in granular HTSC.

We have studied here the magnetoresistance in granular Bi(Pb)-HTSC with trapped magnetic field and for the first time observed and explained the negative magnetoresistance phenomenon.

## 2. Experimental

Ceramic samples of nominal composition $Pb_{0.5}Bi_2Sr_3Ca_4Cu_5O_{16}$ were used for the magnetoresistance study. The samples were pellet-shaped and had diameter of 0.7 cm and height of 0.3 cm.

The usual four – point method was used for resistive measurements.

Samples have the upper temperature and width of resistive transition $T_c$ = 108 -110 K and $\Delta T_c$ = 10 –12 K.

In investigated ceramics the value of $\Delta T_c$ depends strongly on current and magnetic field. The rise of transport current density up to 0.1 A/cm$^2$ and magnetic field up to 20 Oe increases $\Delta T_c$ up to 30 K and as a result the final temperature of superconducting transition temperature drops to liquid nitrogen temperature 77.4 K.

The width of transition interval $\Delta T_c$ increases also as a result of magnetic flux trapping and appearance of frozen magnetoresistance.

The MFT was realized in zero field cooled regime by magnetic pulse with amplitude of $H_i$ = 10 - 200 Oe at T = 77.4 K. Then the magnetic field and angular dependences of magnetoresistance $\Delta R$ (H, α) ( α - angle between the directions of the field H and pulse field $H_i$) were measured for various trapping inducing field $H_i$, transport current and their orientation relative to H.

## 3. Results

It is found that for $\mathbf{H} \parallel \mathbf{H_i} \perp \mathbf{J}$ the field dependence of magnetoresistance of Bi-ceramics with trapped magnetic field, $\Delta R(H) = R(H) – R(0)$, is nonmonotonic and at that $\Delta R$ is negative for small fields $H < H_{inv}$ and changes sign to positive at $H = H_{inv}$.

The normalized field dependences of magnetoresistance $\Delta R(H)$ for $\mathbf{H} \parallel \mathbf{H_i}$ are shown on fig. 1.

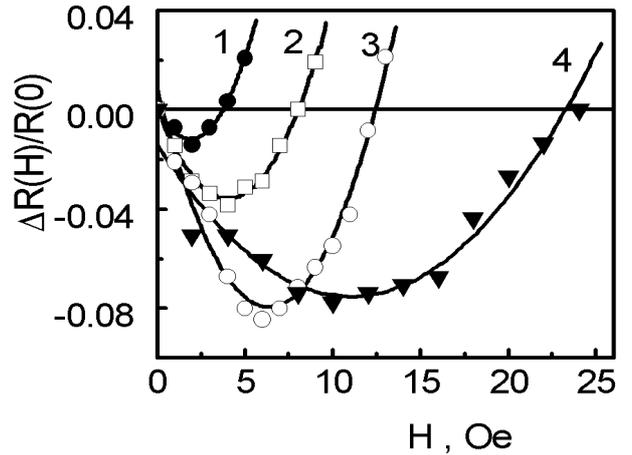

Fig.1. Field dependence of magnetoresistance for various MFT inducing field at T = 77.4 K.
1 - $H_i$ = 80 Oe ;   2 - $H_i$ = 100 Oe;
3 - $H_i$ = 130 Oe;   4 - $H_i$ = 200 Oe

One can see that the phenomenon of negative magnetoresistance (NMR) becomes more strongly pronounced with increase of $H_i$ and hence trapped magnetic field.

In a resistive transition region of granular HTSC the magnetic fields trapped in superconducting clusters are closed through normal regions of the HTSC. As a result the arising trapped magnetic fields are sign-varying. These sign-varying TMF can be characterized by its effective value $H_{teff}$ which is equal to external magnetic field that give rise to the same resistance as effective TMF [4].

It was found that the magnitude of the field of magnetoresistance sign inversion $H_{inv}$ increases almost linearly with the effective value of trapped magnetic field $H_{teff}$ (fig. 2).

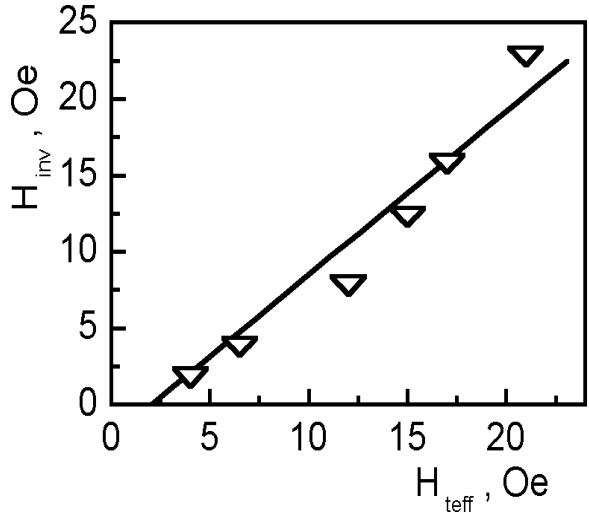

Fig.2. Dependence of magnetoresistance sign inversion field on effective trapped field $H_{inv}(H_{teff})$.

The magnetoresistance field dependences measured for various values of the angle α between the directions of the field H and pulse field $H_i$ are shown on Fig. 3.

It is essential, that the field of inversion of magnetoresistance sign $H_{inv}$ and maximum value of NMR decrease when angle α is increasing and phenomenon of negative magnetoresistance disappears at α → π/2.

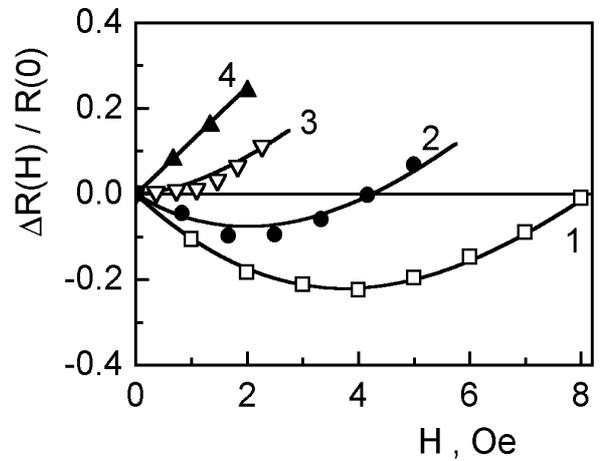

Fig.3. Normalized field dependences of magnetoresistance of HTSC samples with TMF for various values of angle α between directions of the field H and pulse field $H_i$.

1 – α = 0;   2 – α = π/4 ;
3 – α = π/2 ;   4 – α = π.

The angle dependences of magnetoresistance for various values of magnetic field H are presented on fig.4. Note that the increase of H results in decrease of critical angle for magnetoresistance sign inversion.

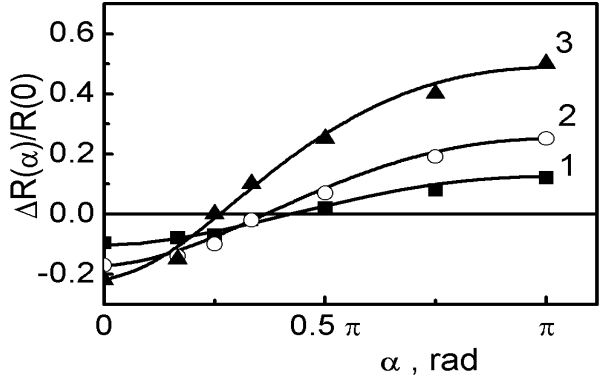

Fig 4. Normalized angle dependences of magnetoresistance of HTSC samples with TMF for various fields H for $H_{teff}$ = 6 Oe and T = 77.4 K.

1 - $H_{teff}$ = 1 Oe ; 2 - $H_{teff}$ = 2 Oe; 3 - $H_{teff}$ = 4 Oe ;

Let us pay attention that the negative magnetoresistance is practically absent for **J** ∥ **H** ∥ **H**$_i$ .

It was found that the field of magnetoresistance sign inversion $H_{inv}$ and maximum value of negative magnetoresistance weakly decreases at grows of a transport current (approximately on 15% at j = 0.3 A/cm$^2$).

The following essential observations should be noted.

1. Rather small fields H are used for magnetoresistive measurements that don't effect the trapped magnetic fields and for the reason all obtained results are quite reproducible.

2. The NMR was also observed in magnetronic films of Bi – HTSC and is obviously the characteristic feature of granular HTSC.

### 4. Discussion.

It is well known that HTSC ceramics have strongly heterogeneous granular structure in which individual superconductor granules are connected through weak links. The concept of such granular system as a percolation Josephson medium with a wide spread of Josephson junctions parameters is generally accepted at present [2].

The resistive properties of the medium are determined by weak links of conductivity channels. In particular, the magnetoresistance is due to destruction of weak links superconductivity by magnetic field.

Magnetic flux trapping in the resistive state of the Josephson medium is realized in HTSC grains (granules) or in grains clusters including weak links (superconducting loops).

The arising local trapped magnetic fields are highly inhomogeneous and sign–varying [4] and so they cannot be observed by usual methods (Josephson interferometers, Hall probes). At the same time such sign-varying TMF destroy the superconductivity of weak links of the conductivity channels and so result in a well known phenomenon – frozen magnetoresistance.

Near the weak links of current channels the directions of trapped magnetic fields are mainly opposite to field inducing the trapping .

So the increase of external field H, on the one hand bring about the decrease of local resulting fields $H_r$ in the regions with high trapped fields $H_t$ ( $H_r = H_t - H$ for $H_t > H$) and on the other hand – the growth of $H_r$ in regions with low trapped fields ( $H_r = H - H_t$ for $H > H_t$). The former results in transition of weak links with critical fields $H_c > H_t - H$ to superconducting state, while the latter factor leads to transition of weak links with $H_c < H - H_t$ to normal state.

For high trapped magnetic fields the first factor prevails in weak fields H and results in network resistance drop and in NMR, the second one becomes dominant for rather high fields H and gives rise to growth of resistance and to inversion of magnetoresistance sign at $H \sim H_{teff}$. So in accordance with experimental results the interplay of these tendencies leads to non-monotone field dependence of system magnetoresistance.

In conclusion, the NMR effect in granular Bi-HTSC was observed, studied and explained on the basis of notions of Josephson weak links medium and local magnetic flux trapping in the medium.